\documentclass[12pt,a4paper]{article}

\usepackage[utf8]{inputenc}
\usepackage[T1]{fontenc}
\usepackage[english]{babel}
\usepackage{amsmath, amssymb}
\usepackage{graphicx}
\usepackage{hyperref}
\usepackage{geometry}
\usepackage{graphicx}

\usepackage{mathrsfs} 
\usepackage{authblk}

\usepackage{caption}       
\usepackage{subcaption}    
\usepackage{geometry}      

\usepackage{comment}
\usepackage{titlesec}

\usepackage{booktabs}
\usepackage{multirow}
\usepackage{pdflscape}

\graphicspath{{Fig/}}
\usepackage{float}

\title{\textbf{Black Holes Thermodynamic Topology in Sharma-Mittal Statistics }}

\author{Issam Najma\thanks{Email: issam.najma.d25@ump.ac.ma}$^{1}$, Yahya Ladghami\thanks{Email: yahya.ladghami@ump.ac.ma}$^{1}$, 
Amine Bouali\thanks{Email: a1.bouali@ump.ac.ma}$^{1,2,3}$, and Taoufik Ouali\thanks{Email: t.ouali@ump.ac.ma}$^{1,2}$
}
\affil{$^{1}$Laboratory of Physics of Matter and Radiation, Mohammed I University, BP 717, Oujda, Morocco}
\affil{$^{2}$Astrophysical and Cosmological Center, BP 717, Oujda, Morocco}
\affil{$^{3}$Higher School of Education and Training, Mohammed I University, BP 717, Oujda, Morocco}

\date{\today}

\begin{document}

\maketitle

\begin{abstract}
In this study, we explore the thermodynamic topology of black holes within the Sharma-Mittal entropy framework. Our study covers several black hole solutions, including charged and uncharged black holes in $d$-dimensional, as well as Schwarzschild and Reissner-Nordström black holes. By introducing the Sharma-Mittal entropy characterized by two parameters $\delta$ and $R$, we explore how deviations from the standard Boltzmann-Gibbs statistics modify the thermodynamic structure and stability properties of these systems. Using Duan's $\Phi$-mapping theory, we compute the corresponding topological numbers and classify black holes into distinct topological categories according to the winding number $W$. Moreover, our analysis shows that, within the Sharma-Mittal formalism, the topological classification is independent of spacetime dimension in the case $d>4$, although dimensions higher than four exhibit distinct features compared to the four-dimensional Schwarzschild and Reissner-Nordström black holes, reveal different behavior of thermodynamic topology under generalized statistics. 
\end{abstract}
\section{Introduction}
\hspace*{0.6cm}Black holes occupy a central position in modern theoretical physics and astrophysics, providing a unique arena for testing our understanding of gravity under extreme conditions. As exact solutions of Einstein's field equations in general relativity, they are characterized by the presence of an event horizon beyond which neither matter nor light can escape. Beyond their classical description, black holes have become indispensable laboratories for investigating quantum gravity and the possible unification of general relativity with quantum mechanics. The discovery of gravitational waves by the LIGO/Virgo Collaboration in 2016 and the first direct image of a black hole by the Event Horizon Telescope in 2019 have transformed black hole physics into a precision observational science \cite{Abbott2016}. These groundbreaking achievements have not only confirmed key predictions of general relativity but have also opened unprecedented opportunities to probe the fundamental nature of spacetime, strong-field gravity, and the quantum structure of the Universe.

Beyond their geometric description, black holes exhibit remarkable thermodynamic properties. In particular, the entropy of a black hole is proportional to the area of its event horizon \cite{Bekenstein1973}, establishing a profound connection between gravity, thermodynamics, and quantum theory. The seminal contributions  of Stephen Hawking and Jacob Bekenstein revealed a fundamental analogy between the laws governing black holes and those of classical thermodynamics \cite{Bardeen1973}.

In this context, the discovery of thermal radiation, known as Hawking radiation, marked a major breakthrough by showing that black holes possess a temperature, referred to as the Hawking temperature \cite{Hawking1975}. This result has profoundly reshaped our understanding of these objects and has opened the way to the study of phenomena such as phase transitions, critical phenomena, and the stability of gravitational systems.

To further deepen the understanding of black hole physics, Wei, Liu, and Robert B. Mann introduced a novel framework that incorporates topology into black hole thermodynamics by interpreting black holes as topological defects in the thermodynamic parameter space \cite{Wei2022}.

This approach relies on the generalized off-shell free energy together with Duan’s topological current theory known as $\Phi$-mapping method \cite{Wei2022, Duan1976}, enabling the characterization of both local and global thermodynamic stability through topological invariants such as winding numbers and topological charges. In this context, a positive winding number signals a locally stable black hole, whereas a negative value corresponds to instability. Based on these topological charges, black holes can be systematically classified into distinct categories-typically three, although a possible fourth class has been suggested \cite{Wei20244}. An alternative formulation using the residue method has also been proposed, yielding results consistent with those obtained via Duan’s theory. Owing to its simplicity, robustness, and conceptual clarity, this topological framework has rapidly gained attention as an effective tool for analyzing thermodynamic stability, phase transitions, and critical phenomena \cite{Wei2020, Wu2023}. It has been successfully applied to a wide range of scenarios, including black holes in anti-de Sitter spacetime, the study of the Hawking-Page phase transition and its holographic interpretation as a confinement-deconfinement transition, as well as investigations incorporating quantum gravity corrections such as higher-derivative terms in Einstein-Gauss-Bonnet and Lovelock theories \cite{Bai2023, Rathi2024, HawkingPage1983}. Furthermore, while initially developed for static configurations, this topological method has been extended to rotating black holes, providing deeper insights into their thermodynamic behavior, stability structure, topological classification, and even properties such as topological photon spheres \cite{Wu2023, HawkingPage1983, Wei2022b}.

In this context, black holes thermodynamic can be generalized by employing statistical entropy forms with multiple parameters, which allow one to go beyond the limitations of classical extensivity. The Sharma-Mittal entropy therefore provides a natural and relevant extension for investigating the thermodynamic properties of black holes \cite{Masi2005, Ghaffari, Ladghami2024e}.

In the present work, we combine topological methods with Sharma-Mittal statistics to investigate the thermodynamic topology of black holes. This approach enables a precise thermodynamic analysis at both local and global thermodynamic stability within the framework of generalized statistics. It reveals the impact of the general form of entropy on the thermodynamic behavior of black holes, as well as on their associated topological properties.

Furthermore, it is well known that Schwarzschild black holes are thermodynamically unstable within the standard Gibbs-Boltzmann formalism \cite{Czinner2016}, making them a useful prototype for probing the effects of alternative statistical descriptions. As we will demonstrate, within the Sharma-Mittal framework, Schwarzschild black holes can exhibit both stable and unstable thermodynamic phases. On the other hand, for the Reissner-Nordström black hole within the standard Gibbs-Boltzmann, the off-shell free energy landscape reveals an unstable small black hole with $w = -1$ and a stable large black hole with $w = +1$. These opposing winding numbers yield a total topological invariant of $W = 0$, providing a robust topological classification of its thermodynamic stability \cite{Wei2022}.

In this paper, we study several classes of $d$-dimensional black hole solutions, including charged and uncharged, in order to analyze the impact of Sharma-Mittal statistics on their stability and other properties. In the same context, particular attention is given to the role of electric charge and spacetime dimensionality within this generalized framework.

The paper is organized as follows. In Section II, we review the Sharma-Mittal entropy in thermodynamics and topological methods. In section III, we apply this approach to various black hole solutions in $d$-dimensions, with particular emphasis on charged and uncharged black holes. Finally, in the  section IV, we present our discussion and conclusions. 

\section{ Sharma-Mittal entropy and topological thermodynamics}
\hspace*{0.6cm}In this section, we present a brief review of the Sharma-Mittal entropy. Additionally, we present the thermodynamic topology of black holes within this framework. The Sharma-Mittal  entropy, a generalized form of entropy that incorporates both Rényi and Tsallis entropies, is given by \cite{Masi2005, Gashti2023}
\begin{equation}
S_{SM} = \frac{1}{R} \left( (1 + \delta S_T)^{\frac{R}{\delta}} - 1 \right), 
\end{equation}
where $S_T$ represents the Tsallis entropy, $R$ and $\delta$ are non-extensivity parameters. In the particular formalism where $\delta=R=1$, the Bekenstein Hawking entropy is recovered \cite{Ladghami2023, Jahromi2018, Rani2024}. Furthermore, in the limiting cases, when $R \to 0$, the Sharma-Mittal entropy reduces to the Rényi entropy, while in the appropriate limit of $\delta$, it reduces to the Tsallis entropy.\\

The black hole temperature within the Sharma-Mittal statistical framework is given by

\begin{equation}
T_{SM} = \left( \frac{\partial S_{SM}}{\partial M} \right)^{-1},
\end{equation}
where M is the black hole mass. Moreover, a central quantity is the generalized off-shell free energy, which characterizes the systems stability and phase structure. In the Sharma-Mittal framework, it can be written as \cite{Wei2022}

\begin{equation}
\label{eq:off-shell}
\mathcal{F}_{SM} = M - \frac{S_{SM}}{\tau},
\end{equation}
where \( \tau \) is the inverse temperature of the system. When \( \tau = \tau_{SM} = 1/T_{SM} \), the off-shell free energy in Sharma-Mittal statistics is recovered.

\begin{figure}[h]
\centering
\includegraphics[width=0.8\textwidth]{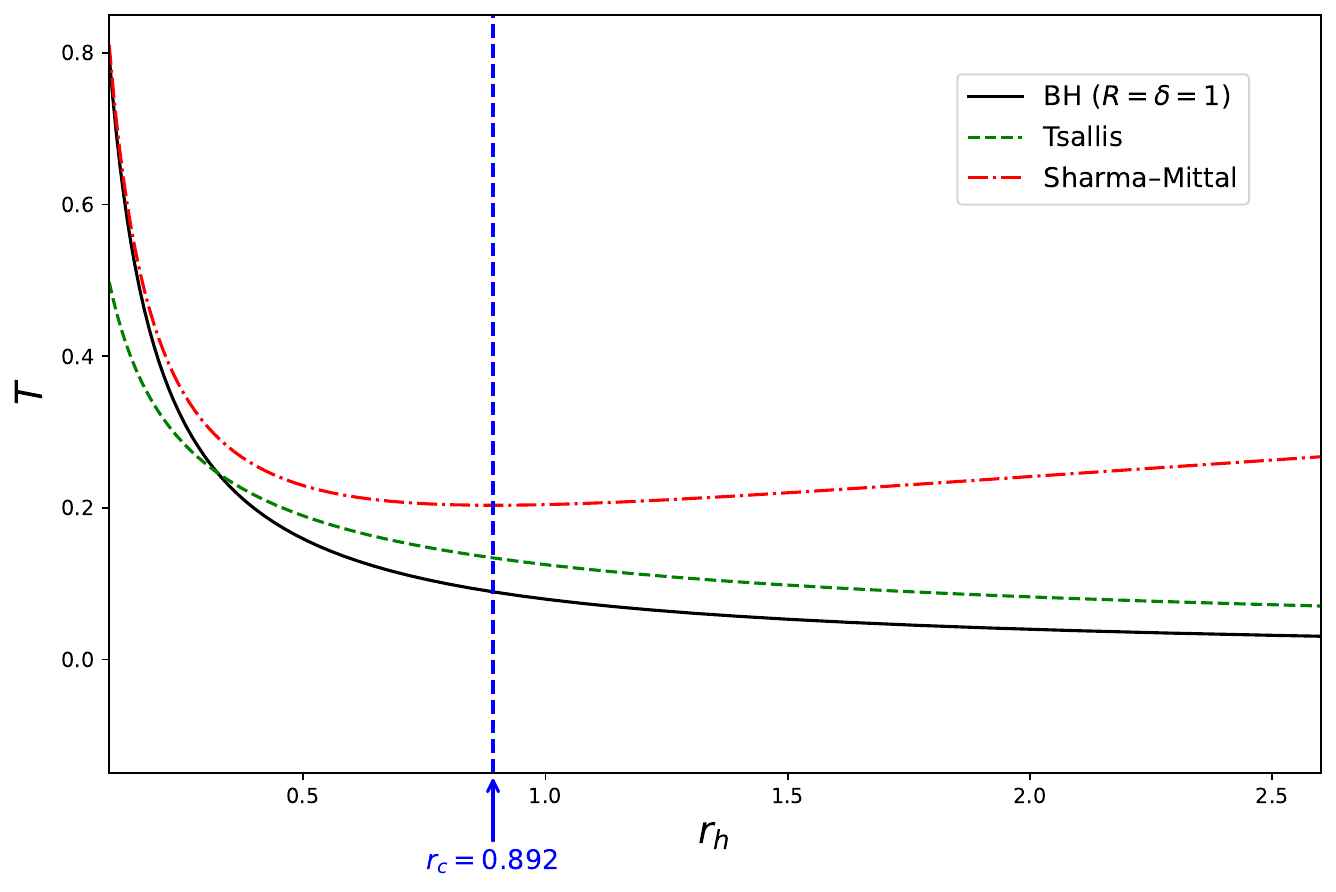}
\caption{The Temperature \(T\) of a Schwarzschild black hole as a function of radius \(r_h\) of Bekenstein-Hawking (BH), Tsallis, and Sharma-Mittal, with parameters  \(R = 0.2\), \(\delta = 0.9\).}
\label{fig:Temp-1a}
\end{figure}

Fig~\ref{fig:Temp-1a} presents the temperature of a Schwarzschild black hole as a function of the horizon radius for three different thermodynamic statistics considered in the literature: Gibbs-Boltzmann statistics, represented by the Bekenstein-Hawking entropy, Tsallis statistics, and Sharma-Mittal statistics.

For small values of $r_h$, all thermodynamic frameworks predict high temperatures, indicating that smaller black holes are thermodynamically hotter, consistent with the standard Hawking radiation behavior. In the Bekenstein--Hawking and Tsallis cases, the temperature decreases monotonically as the horizon radius increases, with the Tsallis curve lying slightly above the Bekenstein--Hawking curve beyond the critical radius $r_{c}=0.892$, due to non-extensive corrections. Conversely, the Sharma-Mittal temperature exhibits a  non-monotonic behavior it initially decreases, reaches a minimum at a critical radius, and then increases again. This indicates the presence of a critical point separating two thermodynamic branches, revealing a more complex stability structure compared to the other entropy frameworks.

A central quantity is the generalized off-shell free energy, which characterizes the system's stability and phase structure. The thermodynamic phase structre and stability are studied from a topological perspictive using Duan's $\Phi$-mapping theory. In this formalism, We define a vector field $\phi$ as follows \cite{Wei2022, Wei2022b, Hazarika2024, Li2022}
\begin{equation}
\phi = (\phi^{1}, \phi^{2}) = 
\left(
\frac{\partial \mathcal{F}_{}}{\partial r_{h}}, -\cot{\theta}\csc{\theta}
\right),
\end{equation}
with $r_{h}$ is the radius of the black hole’s event horizon, and $\theta \in [0,\pi]$ is a parameter constrained.

Furthermore, we introduce the topological current as is defined by the following expression \cite{Duan1984, Alipour2023}
\begin{equation}
j^{\mu} = \frac{1}{2\pi}\,\epsilon^{\mu\nu\rho}\,\epsilon_{ab}\,
\partial_{\nu} n^{a}\,\partial_{\rho} n^{b},
\end{equation}
where $\mu, \nu, \rho = 0,1,2$, $a,b=1,2$, and $\partial_{\nu} = \frac{\partial}{\partial x^{\nu}}$ with $x^{\nu} = (\tau_{SM}, r_{h}, \theta)$. Here, $n=(n^{1},n^{2})$ represents a unit vector defined as
\begin{equation}
n^{a} = \frac{\phi^{a}}{\sqrt{(\phi^{1})^{2} + (\phi^{2})^{2}}}.
\end{equation}

The topological current can be written in terms of the Jacobi tensor and the two-dimensional Laplacian Green function associated with the vector field $\phi(x)$. Using the Dirac delta function $\delta(\phi(x))$, the topological current is expressed as follows \cite{Ladghami2023, Duan1998, Fu2000}
\begin{equation}
j^{\mu} = \delta^{2}(\phi)\,J^{\mu}\!\left(\frac{\phi}{x}\right).
\label{eq:jacobi-current}
\end{equation}

The vector Jacobi is expressed as
\begin{equation}
\epsilon^{ab}\,J^{\mu}\!\left(\frac{\phi}{x}\right)
= \epsilon^{\mu\nu\rho}\,\partial_{\nu}\phi^{a}\,\partial_{\rho}\phi^{b}.
\label{eq:jacobi}
\end{equation}

From Eq.~(\ref{eq:jacobi-current}), we find that the topological current vanishes when $\phi^{a}(x^{i})=0$, where we denote the $i$-th solution by $z_{i}$. It can be shown that the topological current is conserved. The topological number $W$ is constructed by means of this conserved current as follows \cite{Wei2022b}
\begin{equation}
W = \int_{\Sigma} j^{0}\,d^{2}x
= \sum_{i=1}^{N} \beta_{i}\,\eta_{i}
= \sum_{i=1}^{N} w_{i},
\label{eq:top-number}
\end{equation}
where $\beta_{i}$ represents  the Hopf index and counts the number of loops of the vector $\phi^{a}$ as $x^{\mu}$ goes around the zero point $z_{i}$ in the space of $\phi$. Additionally, $\eta_{i}$ is the Brouwer degree, equal to the $\text{sign}\left(J^{0}(\phi/x)_{z_{i}}\right)=\pm1$, and $w_{i}$ represents the winding number for the $i$-th zero point of $\phi$ in the domain $\Sigma$.

The Jacobian vector can be expressed for $\mu=0$ using the following relation \cite{Wei2022}
\begin{equation}
J^{0}\!\left(\frac{\phi}{x}\right)
= \frac{\partial \phi^{1}}{\partial r_{h}}
  \frac{\partial \phi^{2}}{\partial \theta}
- \frac{\partial \phi^{1}}{\partial \theta}
  \frac{\partial \phi^{2}}{\partial r_{h}}
= \frac{\partial^{2}\mathcal{F}}{\partial r_{h}^{2}}
\left(1+\frac{\cos^{2}\theta}{\sin^{3}\theta}\right).
\label{eq:J0}
\end{equation}
For the zero point \(\phi=0\), where the value of \(\theta\) is equal to \(\pi/2\), the Jacobian vector becomes
\begin{equation}
J^{0}\left(\frac{\phi}{x}\right)=\frac{\partial^{2}{\cal F}}{\partial r_{h}^{2 }}.
\end{equation}

Now, we find a simplified expression for the winding number at the zero point as follows

\begin{equation}
w_{i}=\mbox{sign}\left(\left[\frac{\partial^{2}{\cal F}}{\partial r_{h}^{2}} \right]_{z_{i}}\right),
\end{equation}
where \(z_{i}\) represents the solution of \(\phi^{1}=0\).


\section{ Stability of $d$-dimensions Black Holes}
\hspace*{0.6cm}In this section, we investigate the thermodynamic topology of various black hole solutions within the framework of Sharma-Mittal entropy. We begin with charged and non-charged black holes in d>4 dimensions. The case d=4 are also studied but separatly as they exibit differnent behavior regarding the case with $d>4$.
\subsection{Charged and Non-Charged Black Holes in $d>4$ dimensions }
\hspace*{0.6cm}We present and analyze the thermodynamic topology of $d>4$ dimensions black holes within the Sharma-Mittal framework, considering both uncharged and charged black holes. The $d$-dimensional static, spherically symmetric line element is given by
\begin{equation}
ds^2 = -f(r) dt^2 + \frac{dr^2}{f(r)} + r^2 d\Omega^2_{d-2}  ,
\end{equation}
where $d\Omega^2_{d-2}$ denotes the line element on the unit $(d-2)$-sphere. The metric function is given by \cite{Yerra2019}
\begin{equation}
f(r) = 1 - \frac{2\tilde{M}}{r^{d-3}} + \frac{\tilde{Q}^2}{r^{2(d-3)}},
\end{equation}
where 
\begin{equation}
M = \frac{d-2}{8\pi G}\omega_{d-2}\tilde{M},
\end{equation}
and
\begin{equation}
Q = \frac{\sqrt{2(d-2)(d-3)}}{8\pi}\omega_{d-2}\tilde{Q}.
\end{equation}
Here, $M$ represents the Arnowitt-Deser-Misner (ADM) mass of the black hole, $Q$ is the Maxwell field charge, and
\begin{equation}
\omega_{d-2} = \frac{2\pi^{(d-1)/2}}{\Gamma\left(\frac{d-1}{2}\right)}
\end{equation}
denotes the volume of the unit $(d-2)$-sphere. The Gamma function $\Gamma\left(\frac{d-1}{2}\right)$ generalizes the factorial to non-integer values, satisfying $\Gamma(k) = (k-1)!$ for positive integers $k$.

The thermodynamic quantities considered in this section, including the Sharma-Mittal entropy, black hole mass, and Hawking temperature, are expressed in terms of the event horizon radius as
\begin{equation}
S_{SM} = \frac{1}{R}\left[\left(1 + \delta\frac{\omega_{d-2}r_+^{d-2}}{4G}\right)^{\frac{R}{\delta}} - 1\right],
\end{equation}
\begin{equation}
M = \frac{(d-2)\omega_{d-2}}{16\pi G}r_+^{d-3} + \frac{2\pi}{G(d-3)\omega_{d-2}}\frac{Q^2}{r_+^{d-3}}, 
\end{equation}
and
\begin{equation}
T_{SM} = \frac{\partial M}{\partial S_{SM}}
= \left[\frac{d-3}{4\pi r_+}
- \frac{8\pi Q^2}
{(d-2)\omega_{d-2}^2 r_+^{2d-5}}
\right]
\left(1 + \delta\frac{\omega_{d-2}r_+^{d-2}}{4G}
\right)^{1-\frac{R}{\delta}},
\end{equation}
respectively. The Bekenstein-Hawking entropy is
\begin{equation}
S_{BH} = \frac{\omega_{d-2}r_+^{d-2}}{4G},
\end{equation}

Accordingly, the off-shell free energy, Eqs.~(\ref{eq:off-shell}), is given by
\begin{equation}
\begin{aligned}
\mathcal{F}_{SM}(r_+) ={}& \frac{(d-2)\omega_{d-2}}{16\pi G} r_+^{d-3} + \frac{2\pi Q^2}{(d-3)\omega_{d-2} r_+^{d-3}} \\
&- \frac{1}{R\tau} \left[ \left( 1 + \delta \frac{\omega_{d-2} r_+^{d-2}}{4G} \right)^{\frac{R}{\delta}} - 1 \right],
\end{aligned}
\end{equation}
and the thermodynamic vector field $\phi^{1}$ is obtained as
\begin{equation}
\begin{aligned}
\phi^1(r_+) ={}& \frac{(d-2)(d-3)\omega_{d-2}}{16\pi G} r_+^{d-4} \left[ 1 - \frac{32\pi^2 G Q^2}{(d-2)(d-3)\omega_{d-2}^2 r_+^{2d-6}} \right. \\
&\left. - \frac{4\pi r_+}{(d-3)\tau} \left( 1 + \delta \frac{\omega_{d-2} r_+^{d-2}}{4G} \right)^{\frac{R}{\delta}-1} \right].
\end{aligned}
\end{equation}
To simplify the analysis, we parameterize the inverse temperature as follows
\begin{equation}
\tau = \frac{1}{\left[ \frac{d-3}{4\pi r_0} - \frac{8\pi Q^2}{(d-2)\omega_{d-2}^2 r_0^{2d-5}} \right]} \left( 1 + \delta \frac{\omega_{d-2} r_0^{d-2}}{4G} \right)^{\frac{R}{\delta}-1}.
\end{equation}
The thermodynamic equilibrium states correspond to the zeros of \(\phi^1(r_+)\). Each zero \(r_+\), is characterized by a topological charge called winding number

\begin{equation}
w_i = \text{sign} \left[ \frac{\partial \phi^1}{\partial r_+} \bigg|_{r_+ = r_i} \right],
\end{equation}
where \(w_i = +1\) denotes a stable branch, while \(w_i = -1\) corresponds to an unstable branch. The total topological charge of the thermodynamic system is then given by

\begin{equation}
W = \sum_{i=1}^N w_i.
\end{equation}

The total winding number is a topological invariant that remains unchanged under continuous variations of the thermodynamic parameters (\(\delta, R\)), except when pairs of critical points are created or annihilated. Therefore, the evolution of the zeros of \(\phi^1\) provides a topological characterization of the thermodynamic phase structure of Sharma-Mittal black holes.
\begin{table}[h]
\centering
\renewcommand{\arraystretch}{1.8}
\resizebox{\textwidth}{!}{%
\begin{tabular}{|c|cc|cc|cc|cc|cc|}
\hline
& \multicolumn{2}{c|}{$\delta=0.2$}
& \multicolumn{2}{c|}{$\delta=0.5$}
& \multicolumn{2}{c|}{$\delta = 0.67$}
& \multicolumn{2}{c|}{$\delta=1$}
& \multicolumn{2}{c|}{$\delta=1.5$}
\\
\hline

& $r_i$ & $w_i$
& $r_i$ & $w_i$
& $r_i$ & $w_i$
& $r_i$ & $w_i$
& $r_i$ & $w_i$
\\
\hline

6$d$-BH
&
\shortstack{0.563571\\0.881925}
&
\shortstack{+1\\-1}
&
\shortstack{0.558262\\1.6036}
&
\shortstack{+1\\-1}
&
\shortstack{0.556205\\8.61665}
&
\shortstack{1\\-1}
&
\shortstack{0.549694}
&
\shortstack{+1}
&
\shortstack{0.550323 }
&
\shortstack{+1}
\\
\hline

7$d$-BH
&
\shortstack{0.585521\\0.880191}
&
\shortstack{+1\\-1}
&
\shortstack{0.582163\\1.50738}
&
\shortstack{+1\\-1}
&
\shortstack{0.580763\\14.3048}
&
\shortstack{+1\\-1}
&
\shortstack{0.576021}
&
\shortstack{+1}
&
\shortstack{0.576451}
&
\shortstack{+1}
\\
\hline

8$d$-BH
&
\shortstack{0.618324\\0.889355}
&
\shortstack{+1\\-1}
&
\shortstack{0.615743\\1.44468}
&
\shortstack{+1\\-1}
&
\shortstack{0.614748\\9.78173}
&
\shortstack{+1\\-1}
&
\shortstack{0.610697}
&
\shortstack{+1}
&
\shortstack{0.611037}
&
\shortstack{+1}
\\
\hline

$W$
& \multicolumn{2}{c|}{0}
& \multicolumn{2}{c|}{0}
& \multicolumn{2}{c|}{0}
& \multicolumn{2}{c|}{+1}
& \multicolumn{2}{c|}{+1}
\\
\hline

\end{tabular}%
}
\caption{Zero points, winding numbers, and topological charges for $d$-dimensional black holes in the Sharma--Mittal framework with $Q=0.8$, $R=0.6$ and $r_{0}=3$.}
\label{tab:tab-Q}
\end{table}

Table~\ref{tab:tab-Q} presents the winding numbers and global topological charges for higher-dimensional charged black holes in the Sharma-Mittal entropy framework with the parameters fixed at $R=0.6$, $Q=0.8$, and $r_{0}=3$. Two distinct topological classes are identified depending on the value of the Sharma-Mittal parameter $\delta$. For all dimensions considered, when $\delta \leq 0.67$, two zero points appear with winding numbers $+1$ and $-1$, resulting in a vanishing global topological charge, $W=0$. This indicates that the contributions of the two topological defects cancel each other. In contrast, for $\delta>0.67$, only a single zero point with a positive winding number remains, yielding a global topological charge $W=+1$. This behavior shows that the disappearance of the negative winding number changes the topological structure, leaving a single locally thermodynamically stable black hole branch in all higher dimensions considered.

To illustarte this case of higher diemnsions, we consider the five-dimensional spacetime (\( d = 5 \)) and in units where \( G = 1 \), the radial component of the field vector and inverse temperature are given by
\begin{equation}
\phi^1 =
\frac{3\pi r_+}{4}
- \frac{Q^2}{\pi r_+^3}
-
\frac{r_+^2}{r_0^2}
\left(
\frac{3\pi r_0}{4} - \frac{Q^2}{\pi r_0^3}
\right)
\left(
\frac{
1+\delta\frac{\pi^2}{2}r_+^3
}{
1+\delta\frac{\pi^2}{2}r_0^3
}
\right)^{\frac{R}{\delta}-1},
\end{equation} 
and
\begin{equation}
\tau =
\frac{3\pi^2 r_0^2}{2}
\frac{
\left(
1 + \delta \frac{\pi^2}{2} r_0^3
\right)^{\frac{R}{\delta}-1}
}{
\frac{3\pi r_0}{4} - \frac{Q^2}{\pi r_0^3}
},
\end{equation}
where $r_{0}$ is an arbitrary parameter with the dimension of length. By solving the equation corresponding to the vector field component, $\phi^{1}=0$, we obtain a unique point located at $r_{+}=r_{0}$ and $\theta=\frac{\pi}{2}$. This point coincides with the black-hole horizon at $r_{+}=r_{0}$, with the inverse temperature evaluated at this horizon radius.

\begin{figure}[H]
\centering
\begin{subfigure}{0.48\textwidth}
\centering
\includegraphics[width=\linewidth]{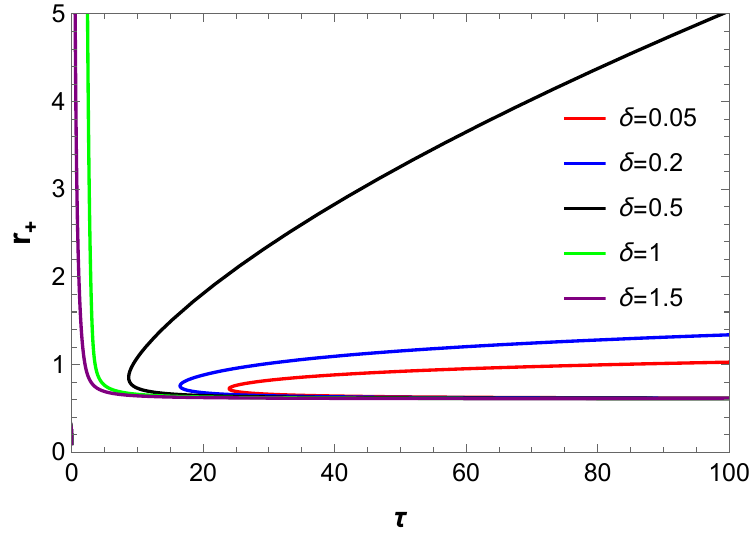}
\caption{Charged black holes}
\end{subfigure}
\hfill
\begin{subfigure}{0.48\textwidth}
\centering
\includegraphics[width=\linewidth]{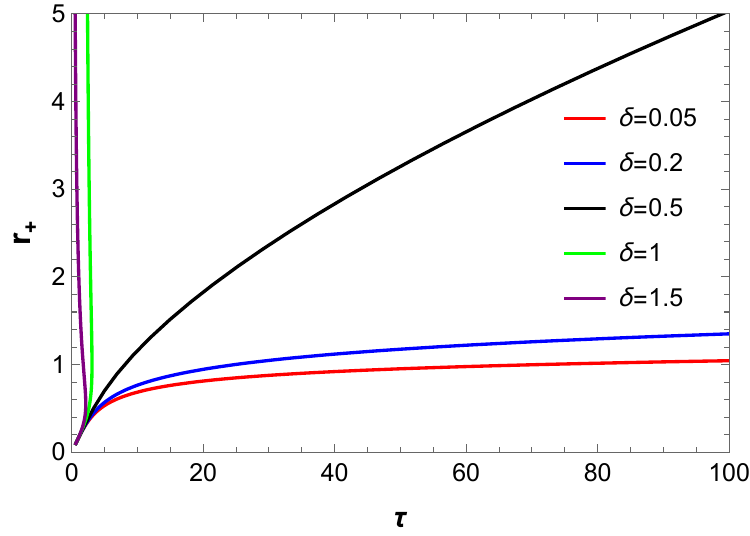}
\caption{Non-charged black holes}
\end{subfigure}
\caption{Zero point curves of $\phi^{1}$ shown in the $\tau-r_{+}$ plane for 5d black hole with $R=0.6$. }
\label{fig:trois}
\end{figure}

\begin{figure}[H]
\centering
\includegraphics[width=1\textwidth]{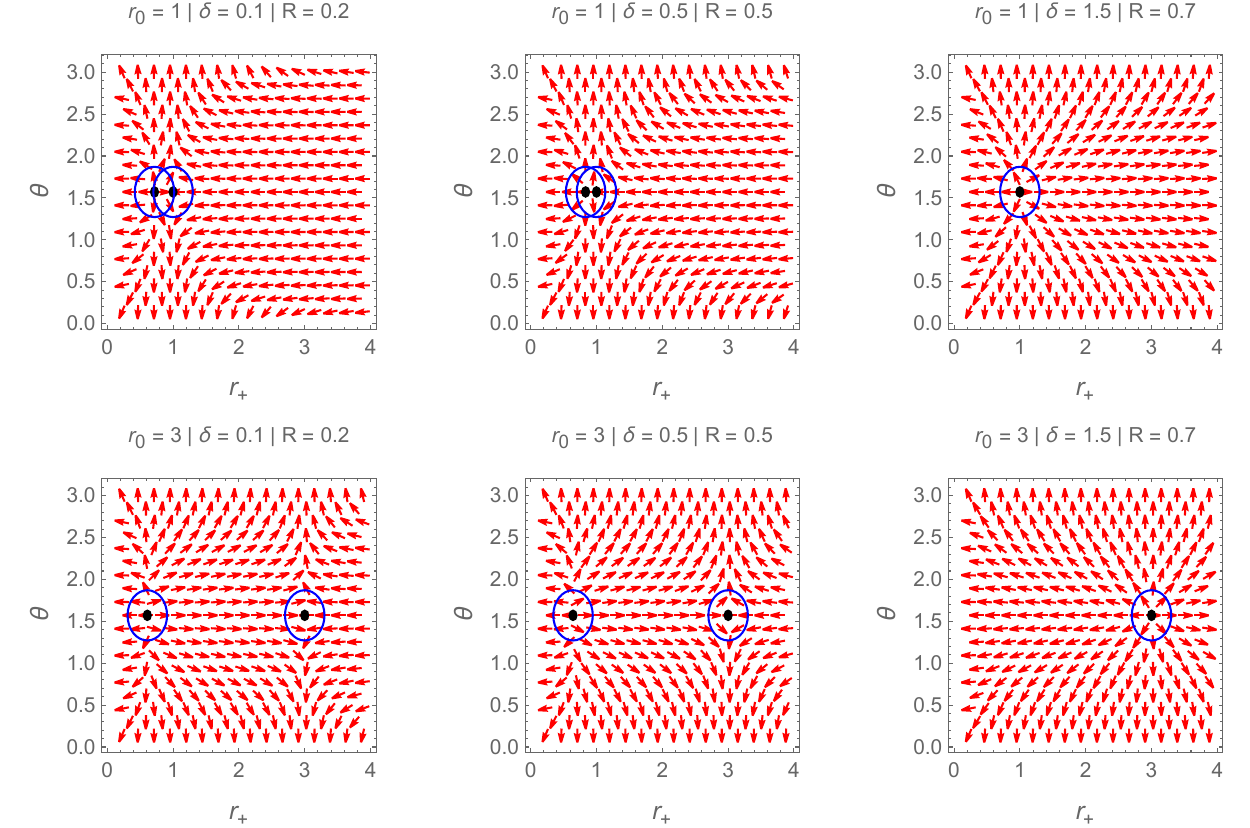}
\caption{Vector field $\phi$ of charged 5d black holes for different values of $\delta $ and $R$  in the
$\theta-r_{+}$ plane.}
\label{fig:charge 5d ph1}
\end{figure}
\begin{figure}[H]
\centering
\includegraphics[width=1\textwidth]{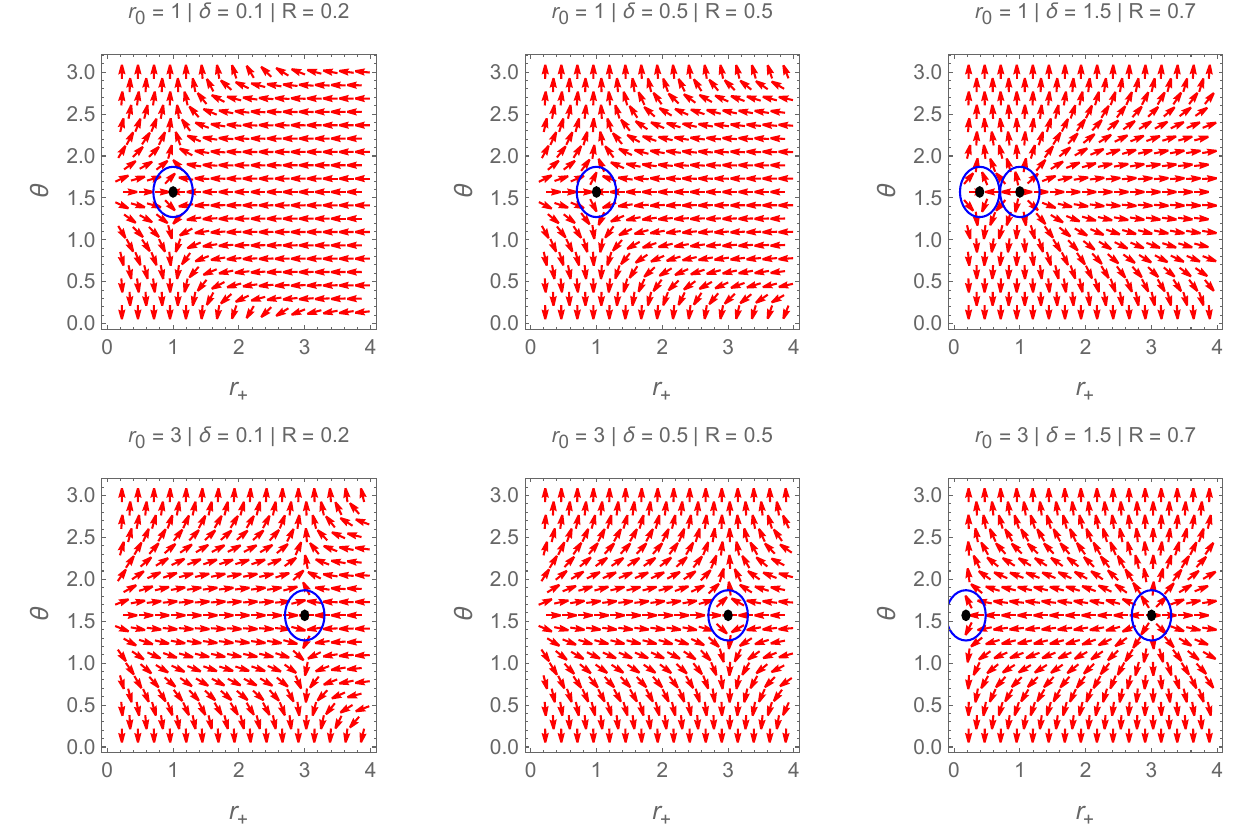}
\caption{Vector field $\phi$ of non charged 5d black holes for different values of $\delta $ ans $R$  in the
$\theta-r_{+}$ plane.}
\label{fig:non-charge 5d ph1}
\end{figure}

In Figs.~\ref{fig:charge 5d ph1}, and~\ref{fig:non-charge 5d ph1}, the topology of the vector field particularly the locations of its zero points demonstrates sensitivity to the parameters $\delta$ and $R$. For the charged case, as shown in Fig.~\ref{fig:charge 5d ph1}, when $\delta < R$, we distinguish two zero points in the $\theta$--$r_+$ plane, which are enclosed by a blue contour. The first zero point is associated with a positive winding number, $w_1 = +1$, indicating a locally stable thermodynamic, whereas the second zero point has a negative winding number $w_2 = -1$, corresponding to a locally unstable thermodynamic. Consequently, the total topological number is $W = w_1 + w_2 = 0$, which gives a globally unstable black hole phase. At the critical value $\delta = R$ ($\delta = 0.5$) and for $r_{0}=0$, the two zero points approach each other and eventually coalesce into a single point. Consequently, the topological number for these cases are moved from  $W = 0$ class to  $W = +1$ class, implying that this class is globally more stable than  $W = 0$ class. For $\delta > R$, as shown in Fig.~\ref{fig:non-charge 5d ph1}, the zero point that appeared earlier splits again into two distinct zero points with opposite winding numbers. Consequently, the total topological number returns to $W = 0$.

In the case $Q = 0$, the topological structure of the vector field $\phi$ behaves differently from the charged scenario, showing reduced complexity due to the absence of electric charge. Nonetheless, the system remains strongly dependent on both parameters $\delta$ and $r_0$.
For $r_0 = 1$ (top panels), when $\delta < R$, the vector field exhibits a single zero point with winding number $w = +1$, indicating a locally and globally stable thermodynamic configuration ($W = +1$). At the critical value $\delta = R$ (here $\delta = 0.5$), the zero point remains stable without any annihilation or creation event, implying no topological phase transition in this regime. However, for \(\delta > R\), two zero points appear, with winding numbers \(w_1 = +1\) and \(w_2 = -1\), respectively. Consequently, the total topological number becomes $W = 0$, signaling the emergence of an unstable phase and a loss of global thermodynamic stability. 

For $r_0 = 3$ (bottom panels), the system initially ($\delta < R$) exhibits two distinct zero points with opposite winding numbers, $w_1 = +1$ and $w_2 = -1$, resulting in a total topological number $W = 0$, indicative of global instability. At $\delta = R$, one zero point vanishes, leaving a single zero point with $w = +1$, which corresponds to a globally stable configuration ($W = +1$). This marks a topological transition from an unstable to a stable phase. For $\delta > R$, another bifurcation occurs: two zero points reappear with opposite winding numbers, restoring $W = 0$. This behavior shows that increasing $\delta$ destabilizes the system once more, reintroducing competing thermodynamic phases.

Overall, in the absence of charge ($Q = 0$), the system exhibits simpler yet still nontrivial topological transitions. The interplay between $\delta$ and $r_0$ governs the creation, annihilation, and bifurcation of zero points, ultimately determining the global thermodynamic stability.


\subsection{Schwarzschild Black Holes}
\hspace*{0.6cm}The case of Schwarzschild Black holes is very interesting as it differ from Both case the  standard Gibbs-Boltzmann thermodynamic and d>4 dimensional spacetime as we will see in this section. Hence, we investigate the thermodynamic topology of Schwarzschild black holes within the framework of Sharma-Mittal statistics. The Schwarzschild black hole metric is
\begin{equation}
ds^{2}=-\left(1-\frac{2M}{r}\right)dt^{2}+\left(1-\frac{2M}{r}\right)^{-1}dr^{2}+ r^{2}d\theta^{2}+r^{2}\sin^{2}\theta d\phi^{2},
\end{equation} 
where $M$ is the mass of the black hole. The event horizon of a Schwarzschild black hole is located at $r_h = 2M$. The Sharma-Mittal entropy, The inverse temperature, and The heat capacity of the Schwarzschild black hole are generalized and are given, by \cite{Ladghami2023}
\begin{equation}
S_{SM}(M) = \frac{1}{R} \left( \left( 1 + 4\delta\pi M^2 \right)^{\frac{R}{\delta}} - 1 \right),
\end{equation}
\begin{equation}
\frac{1}{T_{SM}} = 8\pi M(1 + 4\delta\pi M^2)^{\frac{R}{\delta} - 1},
\label{eq:inv-T}
\end{equation}
and
\begin{equation}
C_v = -\frac{8\pi M^2(1 + 4\delta\pi M^2)^{\frac{R}{\delta}}}{1 + 8\pi  R M^2 - 4\delta\pi M^2}.
\end{equation}
respectively. Using Eq.~(\ref{eq:inv-T}), we obtain
\begin{equation}
T_{SM}(r_{h}) = \frac{1}{4\pi r_{h}}\left(1 + \delta\pi r_{h}^2\right)^{1 - \frac{R}{\delta}},
\end{equation}
To investigate the thermodynamic behavior and determine the topological invariants, we calculate the off-shell free energy, which is given by
\begin{equation}
\mathcal{F}_{SM}=\frac{r_{h}}{2}-\frac{1}{R\tau} \left( \left( 1 + \delta\pi r_{h}^2 \right)^{\frac{R}{\delta}} - 1 \right).
\end{equation}

Furthermore, the components of the vector field can be expressed as
\begin{equation}
\phi^{1}=\frac{\partial\mathcal{F}_{SM}}{\partial r_{h}}= \frac12 - \frac{2\pi r_h}{\tau} \left( 1 + \delta\pi r_h^2 \right)^{\frac{R}{\delta} - 1},
\end{equation}

and
\begin{equation}
\phi^{2}=-\cot\theta\csc\theta.
\end{equation}
We parameterize the inverse temperature $\tau$, in terms of the Sharma-Mittal parameters as follows
\begin{equation}
\tau = 4\pi r_0 \left( 1 + \delta \pi r_0^2 \right)^{\frac{R}{\delta} - 1},
\end{equation}
where $r_{0}$, is an arbitrary positive parameter having the dimensions of length.
Substituting the expression for $\tau$ into $\phi^{1}$, we obtain

\begin{equation}
\phi^1 = \frac{1}{2} - \frac{r_h}{2 r_0}
\left(\frac{1 + \delta \pi r_h^2} {1 + \delta \pi r_0^2}\right)^{\frac{R}{\delta} - 1}.
\end{equation}
The zero-point condition for the vector field \(\phi\) is satisfied when \(\theta = \frac{\pi}{2}\) and \(r_h = r_0\).

The winding number is given by
\begin{equation}
w = \text{Sign} \left[ -1 + \pi (\delta - 2R) r_0^2 \right].
\end{equation}

This shows that the winding number depends only on the combination ($\delta - 2R$). Therefore, the analysis of non-extensive statistical effects on the thermodynamic topology of Schwarzschild black holes provides a natural generalization within the Sharma-Mittal entropy.

\begin{figure}[h]
\begin{subfigure}{0.32\textwidth}
\centering
\includegraphics[width=\linewidth]{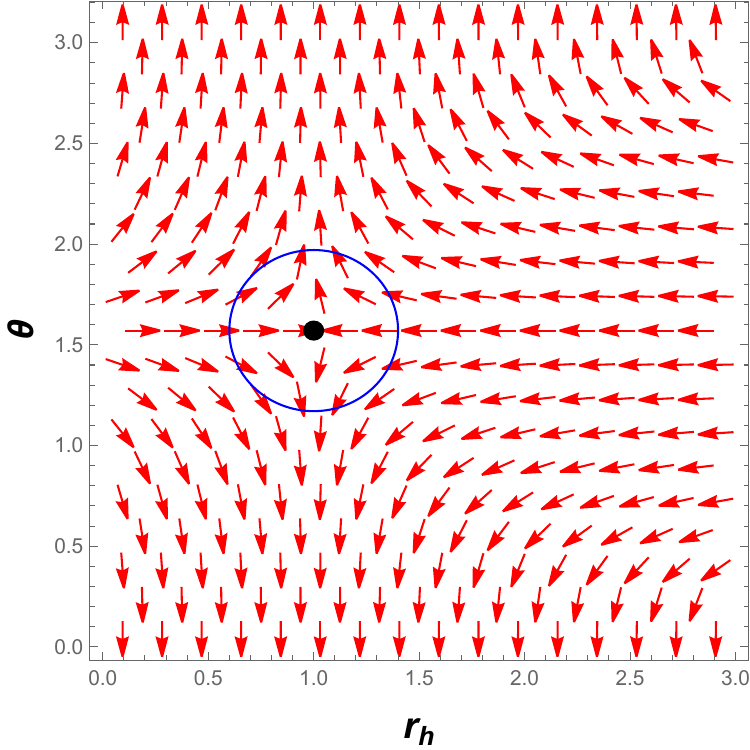}
\caption{$\delta < 2R + 1/\pi r_0^2$}
\end{subfigure}
\hfill
\begin{subfigure}{0.32\textwidth}
\centering
\includegraphics[width=\linewidth]{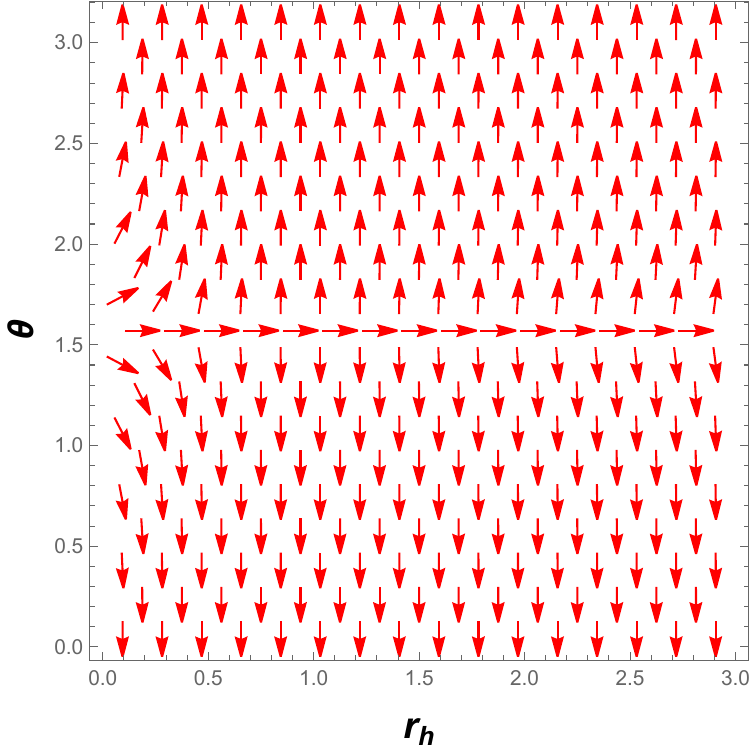}
\caption{$\delta = 2R + 1/\pi r_0^2$}
\end{subfigure}
\hfill
\begin{subfigure}{0.32\textwidth}
\centering
\includegraphics[width=\linewidth]{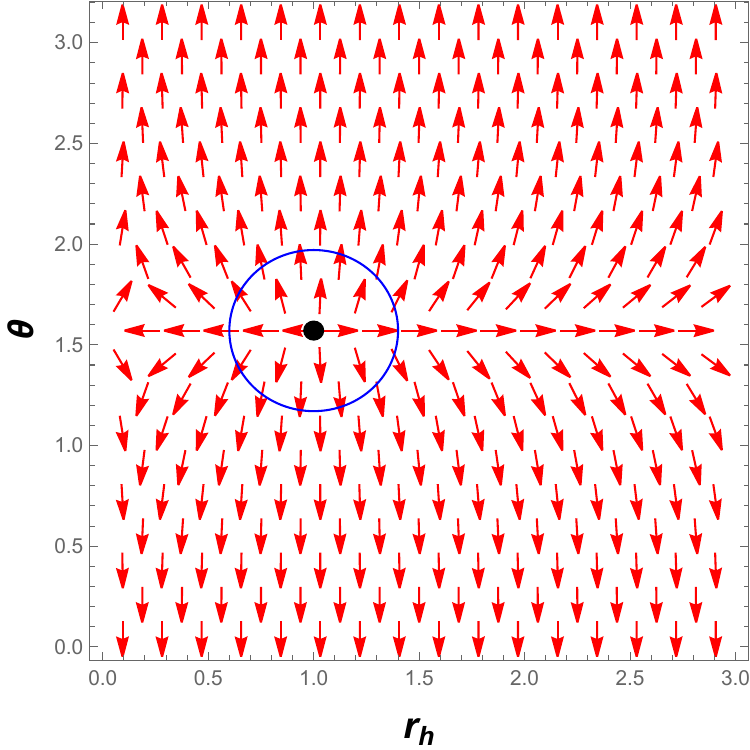}
\caption{$\delta >2R + 1/\pi r_0^2$}
\end{subfigure}

\caption{Vector field $ \phi $ of Schwarzschild black holes for different values of the Sharma-Mittal parameter in
$\theta-r_{h}$ plane.}
\label{fig:phi-sch}
\end{figure}

For $\delta < 2R + 1/(\pi r_0^2)$, the winding number is $w = -1$, for $\delta > 2R + 1/(\pi r_0^2 )$, $w = +1$, 
and for $\delta =2R + 1/(\pi r_{0}^2)$, $w = 0$.
This behavior is directly reflected in the vector field structure shown in Fig~\ref{fig:phi-sch}. For $\delta < 2R + 1/(\pi r_0^2)$, the winding number associated with the zero point of the vector field is $w=-1$. This
configuration indicates the presence of a locally thermodynamically unstable black hole, resulting in a global topological number of $W=-1$. At the critical value $\delta = 2R + 1/(\pi r_0^2)$, no zero points exist and confirming the trivial topological class $W = 0$. For $\delta > 2R + 1/(\pi r_0^2)$, a single zero point emerges with a winding number $w = +1$, corresponding to a globally stable thermodynamic phase with $W = +1$.

\begin{figure}[h]
\centering
\includegraphics[width=0.48\textwidth]{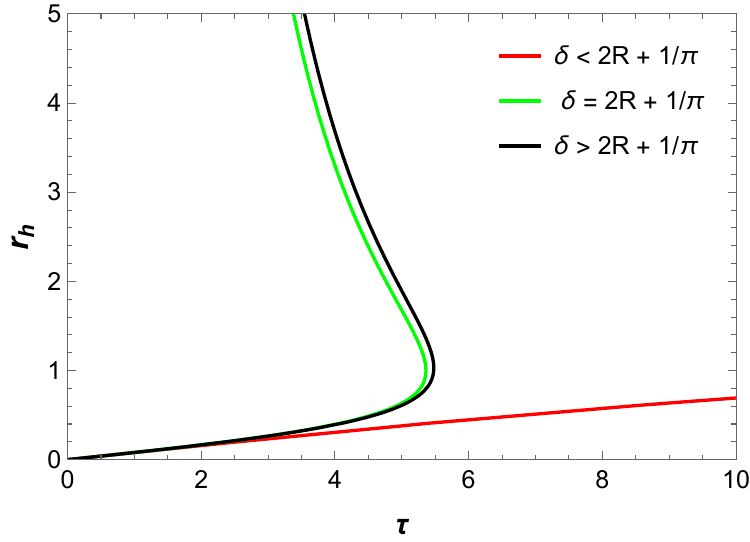}
\caption{ The inverse-temperature function $\tau(r_h)$ for a Schwarzschild black hole in the Sharma--Mittal entropy framework with $R=0.6$ and $r_{0}=1$. }
\label{fig:tau-1}
\end{figure}
  
The thermal evolution of these three classes is presented in Fig~\ref{fig:tau-1}, which shows that two of the classes display  distinct thermal behavior.
For the first class, characterized by $W=-1$ and represented by the red curve, which corresponds to the case where the Sharma-Mittal  parameter exceeds the critical value, we observe that the event horizon increases with the inverse temperature. This behavior indicates that small black holes emit high-temperature Hawking radiation, whereas large black holes radiate at lower temperatures. Such thermodynamic behavior is consistent with that of Schwarzschild black holes in conventional statistical frameworks. Moreover, this class is associated with a negative topological number ($W=-1$), reflecting its thermodynamic instability and remaining consistent with the well-known instability of Schwarzschild black holes in standard descriptions.

The stability of Schwarzschild black holes can further be investigated by studying the evolution of their heat capacity in the framework of Sharma-Mittal statistics which is given by 
\begin{equation}
C_{v} = -\frac{8\pi r_h^2 \left(1 + 4\delta \pi r_h^2 \right)^{\frac{R}{\delta}}}
{1 + 8\pi R r_h^2 - 4\delta \pi r_h^2}.
\end{equation}
\begin{figure}[h]
\centering
\includegraphics[width=0.6\textwidth]{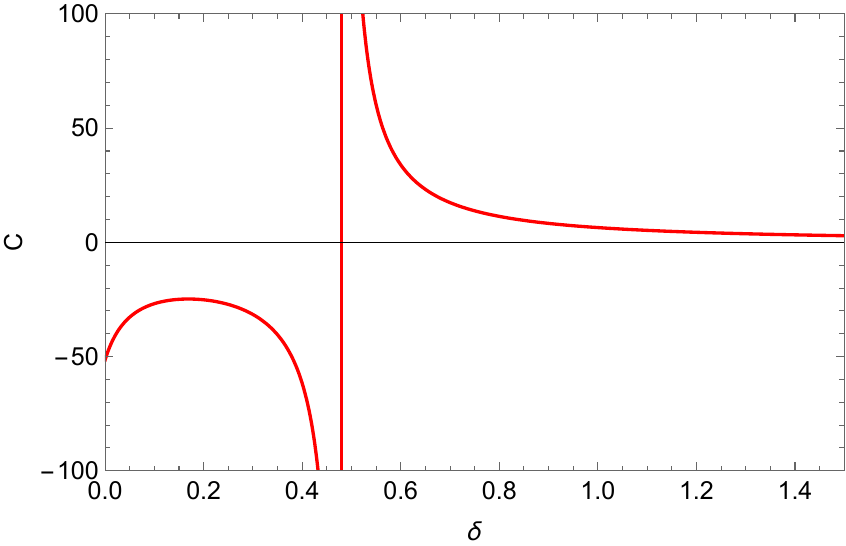}
\caption{Heat capacity of Schwarzschild black hole in the framework of Sharma-Mittal entropy for with $R = 0.2$ and $r_{h}=1$.}
\label{fig:capa-sch}
\end{figure}

The impact of non-extensive entropy on the thermodynamic phase structure of the Schwarzschild black hole can be analyzed through the behavior of the heat capacity, $C_{v}$. Fig~\ref{fig:capa-sch} illustrates the evolution of the heat capacity as a function of $\delta$, with $R$ and $r_h$ held fixed. We observe the existence of two distinct thermodynamic phases. For small values of $\delta$, the heat capacity is negative, indicating that the black hole is thermodynamically unstable. In contrast, for larger values of $\delta$, the heat capacity becomes positive, implying that the black hole is thermodynamically stable.

\subsection{Reissner-Nordström Black Holes}
\hspace*{0.6cm}We investigate the thermodynamic topology of Reissner-Nordström black holes within Sharma-Mittal statistics. The metric function for Reissner-Nordström black holes is  
\begin{equation}
f(r) = 1 - \frac{2M}{r} + \frac{Q^2}{r^2}.
\end{equation}

The mass of Reissner-Nordström black hole in terms of the event horizon \( r_+ \) is obtained by setting \( f(r_+) = 0 \), which yields

\begin{equation}
M = \frac{r_+}{2} + \frac{Q^2}{2r_+}.
\end{equation}

The generalized off-shell free energy of the Reissner-Nordström black hole is 

\begin{equation}
\mathcal{F}_{SM} =  \frac{r_+}{2} + \frac{Q^2}{2r_+} - \frac{1}{R \tau} \left( \left( 1 + 4\delta\pi (r_+)^2 \right)^{\frac{R}{\delta}} - 1 \right).
\end{equation}

The components of the vector field, derived from the off-shell free energy, are
\begin{equation}
\phi^1 = \frac{\partial \mathcal{F}_{SM}}{\partial r_+} =\frac{1}{2}
- \frac{Q^{2}}{2 r_+^{2}}
- \frac{8 \pi r_+}{\tau}
\left(1 + 4 \delta \pi r_+^{2}\right)^{\frac{R}{\delta}-1},
\end{equation}
and
\begin{equation}
\phi^2 = -\cot\theta \csc\theta.
\end{equation}

To simplify our analysis, we parameterize the inverse temperature as follows

\begin{equation}
\tau = \frac{4\pi r_{0}}{1-\frac{Q^{2}}{2 r_{0}^{2}}}\left(1 + 4 \delta \pi r_{0}^{2}\right)^{\frac{R}{\delta}-1},
\end{equation}

\begin{figure}[H]
\centering
\includegraphics[width=0.6\textwidth]{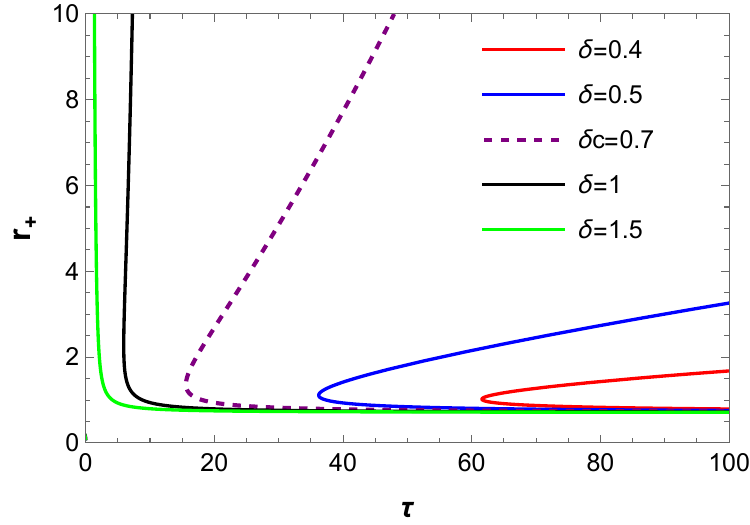}
\caption{Zero-point curves of the vector component $\phi^{1}$ in the $\tau$--$r_+$ plane for different values of $\delta$ parameter, with $R = 0.6$ and $Q = 1$.}
\label{fig:RN-BH-tau-1}
\end{figure}
Fig~\ref{fig:RN-BH-tau-1} shows the thermal evolution of the vector field in the $\tau$-$r_+$ plane for the intermediate values $\delta = 0.4$ and $0.5$. This indicates multiple branches of solutions for a given value of $\tau$ a clear signature of a phase transition, analogous to the behavior observed in black hole thermodynamics. The turning points of the curve correspond to critical points where stability changes, signaling the coexistence of distinct thermodynamic phases.

For larger values of $\delta$ (e.g.\ $\delta = 1$ and $\delta = 1.5$), this feature disappears, and the curves revert to a monotonic profile. In this
regime, the system again exhibits a single thermodynamic branch, indicating that the phase transition seen at intermediate $\delta$
vanishes. This shows that increasing $\delta$ suppresses the phase transition and simplifies the thermodynamic behavior.

Overall, the figure demonstrates that the parameter $\delta$ plays a crucial role in controlling the thermodynamic phase structure. In particular, there exists a critical regime around $\delta \approx \delta _{c}$ where the system undergoes a phase transition, while for both smaller and larger values of $\delta$, the system remains in a single-phase configuration.

\begin{figure}[H]
\centering
\includegraphics[width=1\textwidth]{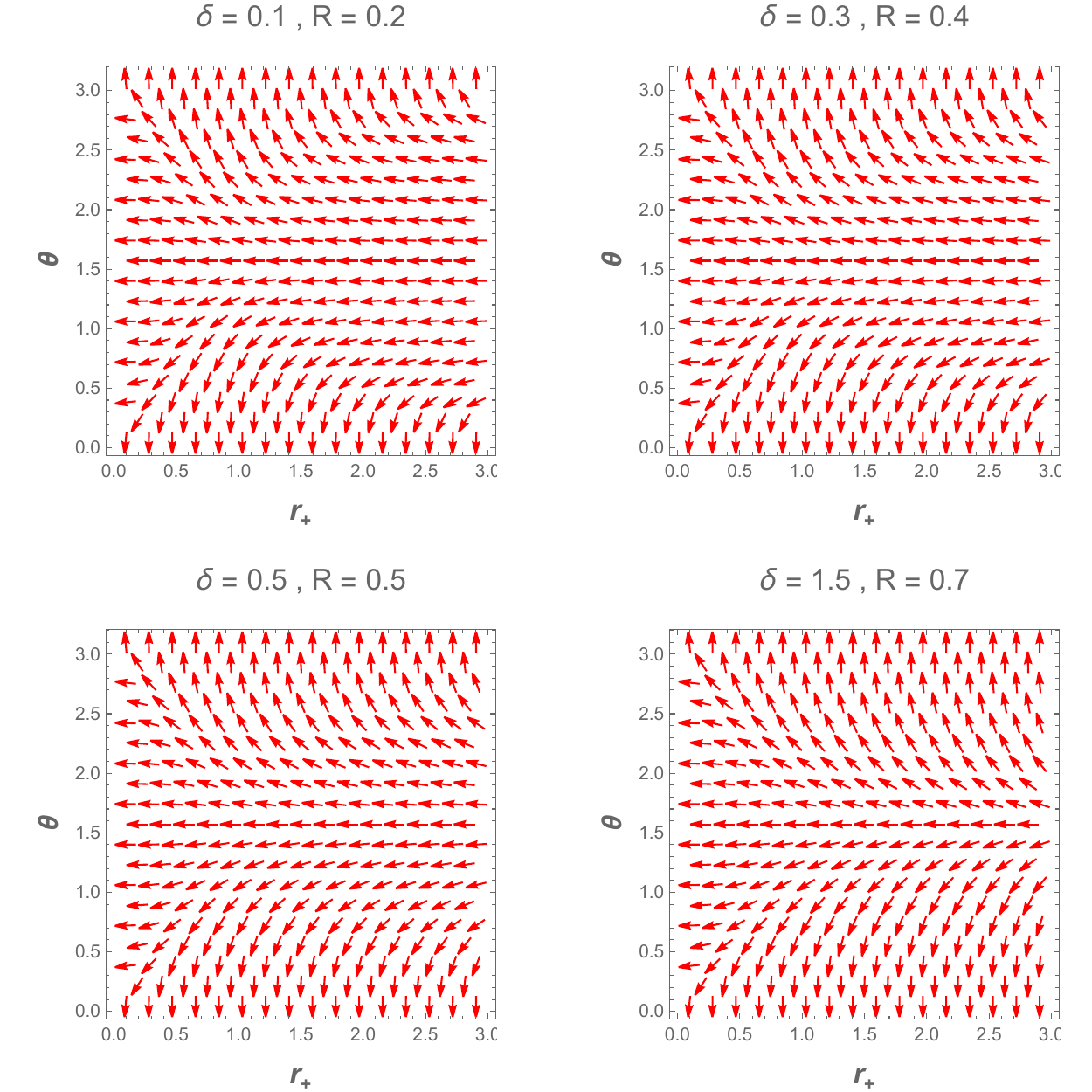}
\caption{Vector field $\phi$ for Reissner-Nordström black holes  in the $\theta$--$r_+$ plane, with $r_0 = 3$ and $Q = 1$.}
\label{fig:RN-BH-Phi-1}
\end{figure}
The vector field structure shown in Fig~\ref{fig:RN-BH-Phi-1} for Reissner-Nordström black holes reveals a qualitatively simple and robust topological behavior. For all considered values of the Sharma-Mittal parameters $(\delta, R)$, no zero points are observed in the  $\theta$--$r_+$ plane. This indicates that the vector field does not exhibit any critical points.

As a consequence, the topological number is $W = 0$ throughout the parameter space. The absence of zero points implies that there are no competing thermodynamic phases, nor any local stability structures characterized by nonzero winding numbers. Therefore, the system does not undergo any topological phase transition as the parameters $\delta$ and $R$ vary.

This behavior indicates that the topological thermodynamics of Reissner-Nordström black holes is insensitive to the Sharma-Mittal entropy deformation. In other words, the non-extensive effects encoded in $(\delta, R)$ do not modify the global topological class of the system. The black hole remains in the trivial class $W = 0$ for all cases considered.

Interestingly, this result is analogous to the behavior observed for several Reissner-Nordström black hole solutions, such as the Reissner-Nordström black hole in the standard Gibbs-Boltzmann thermodynamics, which belong to the same topological class. In particular, this class arises either from the absence of zero points or from the presence of two zero points carrying opposite winding numbers, both of which lead to a net winding number $W=0$. Therefore, the underlying topological structure of the thermodynamic phase space remains unchanged in this framework.




\section*{Conclusions and discussions}

\hspace*{0.6cm}In this paper, we investigated the thermodynamic topology of black holes within the framework of non-extensive statistics. Specifically, we employed the Sharma-Mittal entropy and applied Duan's $\phi$-mapping theory in conjunction with the generalized off-shell free energy. Using this formalism, we analyzed the thermodynamic stability of black holes and characterized their thermodynamic configurations through the winding number and the global topological number. Our analysis was performed for Higher-dimensional charged and uncharged black holes, including five-dimensional, as well as Schwarzschild and Reissner–Nordström black holes. Furthermore, our analysis was carried out in two distinct cases. First, we considered the neutral case by setting $Q=0$. Second, we investigated the charged case while allowing the parameters $\delta$ and $R$ to vary.

For the thermodynamic topology of $d$-dimensional black holes in the Sharma-Mittal entropy framework, two distinct topological classes are identified depending on the value of the Sharma-Mittal parameters. For $\delta \leq 0.67$, $d$-dimensional charged black holes possess two zero points with winding numbers $+1$ and $-1$ for all dimensions considered, yielding a global topological number $W=0$. In contrast, for $\delta>0.67$, $d$-dimensional charged black holes possess a single positive winding number for all dimensions considered, yielding a global topological number $W=+1$. This result indicates that increasing the Sharma-Mittal parameter modifies the thermodynamic topological structure by removing the negative winding number branch and preserving only the locally thermodynamically stable branch. Furthermore, our  analysis reveals that, once $\delta$ exceeds the critical value of approximately $0.67$, the spacetime dimension has only a weak influence on the resulting topological classification.

For five-dimensional charged and neutral black holes, the interplay between the Sharma-Mittal parameters and the reference radius produces rich topological phenomena, including the creation and annihilation of zero points. These bifurcation processes reveal re-entrant topological stability and illustrate how generalized entropy modifies the thermodynamic phase structure through purely topological mechanisms.

For the neutral case, a comparison with the Schwarzschild black hole reveals three distinct thermodynamic classes, depending on the Sharma-Mittal parameters. When
$\delta > 2R + 1/\pi r_0^{2}$,
the global topological number is $W=1$, indicating a thermodynamically stable phase. In contrast, for
$\delta < 2R + 1/\pi r_0^{2}$,
the global topological number becomes $W=-1$, corresponding to a thermodynamically unstable phase, consistent with the standard Schwarzschild black hole. At the critical value
$\delta = 2R + 1/\pi r_0^{2}$,
the global topological number vanishes ($W=0$), marking the transition between the stable and unstable phases. At this critical point, the vector field possesses no zero points.

The Reissner-Nordström black hole exhibits different behavior. Although the thermal evolution shows non-trivial thermodynamic features for intermediate values of $\delta$, the corresponding vector field remains topologically trivial, with no zero points and a global topological number $W = 0$. This behavior is similar to that of many other Reissner-Nordström black holes in the literature. This observation indicates that thermodynamic phase transitions do not necessarily imply changes in the global topological class, highlighting the complementary role of thermodynamic topology.

Overall, the present work demonstrates that thermodynamic topology based on Sharma-Mittal entropy provides a unified framework for investigating black hole thermodynamics beyond the standard Bekenstein--Hawking description. The winding number and the global topological number constitute universal topological invariants that successfully characterize equilibrium states and thermodynamic stability independently of the particular black hole solution. The excellent agreement between the topological analysis and the conventional thermodynamic quantities further supports the effectiveness of this approach.

Finally, the present framework can be extended to a wide range of black hole solutions, including rotating, regular, accelerating, and quantum-corrected black holes, as well as those arising in modified gravity theories. Moreover, the topological invariants introduced in this work, such as the winding number and the global topological number, could serve as physically meaningful features for machine-learning models, enabling efficient classification, stability prediction, and phase-transition detection in black hole thermodynamics. Combining thermodynamic topology with data-driven approaches may therefore provide new insights into the microscopic origin of black hole entropy within generalized statistical mechanics.

\end{document}